\documentclass[11pt]{article}
\usepackage{latexsym} 
\usepackage{graphicx}
\usepackage{amsfonts}
\usepackage{amsmath}
\usepackage{amsthm}
\usepackage{amssymb}

\renewcommand{\theequation}{\thesection.\arabic{equation}}
\newcounter{subequation}[equation]
\makeatletter

\newcommand{\p}{^{\prime}}
\newcommand{\pp}{^{\prime \prime}}
\newcommand{\alphadot}{\stackrel{\cdot}\alpha}
\expandafter\let\expandafter\reset@font\csname reset@font\endcsname

\def\subeqnarray{\arraycolsep1pt
    \def\@eqnnum\stepcounter##1{\stepcounter{subequation}%
        {\reset@font\rm(\theequation\alph{subequation})}}
\jot5mm     \eqnarray}

\makeatother

\def\be{\begin{equation}}

\def\ee{\end{equation}}
\def\bea{\begin{eqnarray}}
\def\eea{\end{eqnarray}}
\def\ba{\begin{array}}
\def\ea{\end{array}}
\def\dd{\partial}

\def\e {\varepsilon}
\def\half{\frac{1}{2}}

\def\one#1{#1^{\raise5pt\hbox{$\scriptstyle\!\!\!\!1$}}\,{}}
\def\two#1{#1^{\raise5pt\hbox{$\scriptstyle\!\!\!\!2$}}\,{}}

\def\II{\hbox{{1}\kern-.25em\hbox{l}}}

\makeatletter
\def\binrel@#1{\begingroup
  \setboxz@h{\thinmuskip0mu
    \medmuskip\m@ne mu\thickmuskip\@ne mu
    \setbox\tw@\hbox{$#1\m@th$}\kern-\wd\tw@
    ${}#1{}\m@th$}%
  \edef\@tempa{\endgroup\let\noexpand\binrel@@
    \ifdim\wdz@<\z@ \mathbin
    \else\ifdim\wdz@>\z@ \mathrel
    \else \relax\fi\fi}%
  \@tempa
}
\let\binrel@@\relax
\def\overset#1#2{\binrel@{#2}%
  \binrel@@{\mathop{\kern\z@#2}\limits^{#1}}}
\def\underset#1#2{\binrel@{#2}%
  \binrel@@{\mathop{\kern\z@#2}\limits_{#1}}}
\makeatother
\newfont{\bbd}{msbm10 scaled\magstep1}

\begin{document}

\vspace*{-1cm}

\begin{center}

{\LARGE { QCD at high energies and Yangian symmetry  }
}\footnote{\it Dedicated to the memory of L.N. Lipatov.}

\vspace{0.5cm}

{\large \sf
R. Kirschner$^a$\footnote{\sc e-mail:Roland.Kirschner@itp.uni-leipzig.de}} 

\vspace{0.5cm}

\begin{itemize}
\item[$^a$]
{\it Institut f\"ur Theoretische
Physik, Universit\"at Leipzig, \\
PF 100 920, D-04009 Leipzig, Germany}
\end{itemize}

\end{center}

\vspace{.5cm}
\begin{abstract}
\noindent
Yangian symmetric correlators provide a tool to investigate 
integrability features of  QCD at high energies. We discuss the
kernel of the equation of perturbative Regge asymptotics, the kernels
of the evolution equation of parton distributions, Born scattering
amplitudes and coupling renormalization.
\end{abstract}

\renewcommand{\refname}{References.}
\renewcommand{\thefootnote}{\arabic{footnote}}
\setcounter{footnote}{0} \setcounter{equation}{0}

\section{Gauge field theories and Quantum integrable systems}
 \setcounter{equation}{0}

In 1993 L.N. Lipatov pointed out the relation of the perturbative QCD 
Regge aymptotics \cite{LNL76,BFKL,BL78} to integrable quantum spin chains 
and showed that the 
contributions of multiple reggeized gluon exchanges can be treated by the
methods of quantum integrable systems \cite{LNL93}. This result has
great significance far beyond the particular questions of Regge asymptotics.  
Up to that time the application of such methods to Quantum field theory was
restricted to 
sophisticated models in $1+1$ dimensions. 

It was clear immediately, that the Bjorken asymptotics \cite{BjP69, GL72a,
GL72b,LNL74,AP77,YLD77}
and the composite
operator renormalization have similar features of integrability \cite{BDM98}. 
In both
cases the Yangian symmetry underlying integrability is based on 
conformal symmetry, which is broken by loop corrections in the case of 
QCD. The breaking is suppressed by supersymmetry and is absent in
$\mathcal{N} = 4 $ super Yang-Mills theory.    
Much work has been devoted in the last two decades to super Yang-Mills
under this aspect.  Quantum integrability works in the composite operator
renormalization in the planar limit to all orders \cite{BS03} and
in the computation of scattering amplitudes and Wilson loops
\cite{DHKS08,DHP09,AH}.

Yangian symmetric correlators (YSC) have been proposed as a convenient
formuation of the Yangian symmetry of amplitudes and are shown to provide a 
tool of construction \cite{CK}. YSC depend on a set of spectral parameters,
some combination of which are related to the helicities of scattering
particles.
The construction method of YSC are related to the method of on-shell graphs
in the amplititude  construction \cite{Broedel}. 
In the original form  of the latter method \cite{AH}
no such parameters appeared. In a number of papers the deformation of
amplitude expressions by  such parameters has been studied
and their eventual role for regularization of loop divergencies has been
discussed \cite{FLMPS13,FLS14,BBR14,BHLY14}.

In the present contribution the notion of YSC is recalled and a few examples
of their construction by $R$ operations are given, which appear in the
following applications.

We explain how the kernel of the BFKL equation  in dipole form
\cite{YB99,YK99,BLV04} 
emerges from a four-point YSC based on $s\ell_2$ symmetry. 
Based on \cite{FK16,KS17} we
discuss the relation of the parton evolution kernels 
and the amplitude of parton splitting \cite{LNL74} to three-point YSC based
on $s\ell_2$ symmetry. 
The relation of the Born level scattering amplitudes of gluons and quarks
to four-point YSC with $s\ell_4$ symmetry is considered. 
Finally, we consider the convolution of two three-point YSC, similar to a
gluon self-energy Feynman graph, and show how the leading Gell-Mann - Low
coefficient of the coupling renormalization appears. This relation is
actually  known in terms of the parton splitting amplitude \cite{LNL74}
and the momentum sum rules for parton evolution kernels  \cite{AP77} 
and has been shown recently to extend to 
arbitrary helicity values \cite{GS14}. 

\section{Yangian symmetric correlators}
\label{YSC1}
 \setcounter{equation}{0}

Consider $n$ Heisenberg canonical pairs, $x_a, \dd_a, a=1, ..., n$,
$$ [\dd_a, x_b] = \delta_{ab}, [x_a, x_b] = 0, [\dd_a, \dd_b] = 0. $$ 
$ L_{ab}= \dd_a x_b $ obey the general linear $g\ell_n$ Lie algebra  
commutation relations. We define  the $L$ operator as a $n \times n$ matrix
with these entries with the unit matrix $I$ multiplied by the
spectral parameter $u$ added,
$$ L(u) =  u I + L(0), L(0)_{ab} = L_{ab}  = \dd_a x_b. $$
This operator acts on functions of $x_a, a=1, ..., n$. We consider the case
of homogeneous  functions with the weight denoted by $2\ell$, e.g.
$ x_n^{2\ell} \cdot \psi(\frac{x_1}{x_n} , ...\frac{x_{n-1}}{x_n}) $.
The $L$ operator restricted to such homogeneous functions depends
additionally on the weight $2\ell$ or on $u^+ = u+2\ell$, 
$$L(u)  x_n^{2\ell} \cdot \psi =  x_n^{2\ell} L(u^+, u) \psi. $$

We need $N$ copies of the above sets of canonical pairs,
$x_{i,a}, \dd_{i,a} $ and of the $L$ operator 
$L_i(u_i, u_i^+), i = 1, ..., N $
to define the monodromy matrix operator in terms of the matrix product

$$ T(\mathbf{u}) = \prod_1^N L_i(u_i^+, u_i) $$
It is convenient to display the parameters in the form

$$ \mathbf{u} = 
\begin{pmatrix}
u_1 & ...& u_N \\
u_1^+ & ...& u_N^+
\end{pmatrix}
$$ 

Further, we consider functions of $N$ points in $n$ dimensional space,
homogeneous with respect to the coordinates of  each point
 $\mathbf{x}_i$  of weight $ 2 \ell_i$. We define a Yangian symmetric
correlator (YSC) 
 to be such a homogeneous function of $N$ points obeying the
eigenvalue relation with the monodromy matrix

\be 
\label{YSC} 
T(\mathbf{u}) \Phi(\mathbf{x}_1,...\mathbf{x}_N; \mathbf{u})  = E(\mathbf{u})
\Phi(\mathbf{x}_1,...\mathbf{x}_N; \mathbf{u}). \ee

The monodromy matrix can be separated into two factors, e.g.
$ T(\mathbf{u}) = T_1(\mathbf{u}_1)T_2(\mathbf{u}_2) $ where 
$T_1$ denotes the product of the first $L_i, i=1, ..., N_1$ and $T_2$
the product of the subsequent $L_i, i= N_1+1, .., N$. 
Then $\Phi = \Phi_1 \Phi_2$ solves (\ref{YSC}) (with $E = E_1 E_2$) if 
$ T_1 \Phi_1 = E_1 (\mathbf{u}_1) \Phi_1$ and  
$ T_2 \Phi_2 = E_2 (\mathbf{u}_1) \Phi_2 $. 
  
The case of one point is trivial, but provides  a convenient starting
point. We have two obvious one-point YSC,
$ L(u,u) \cdot 1 = (u+1) 1,  \ \ L(u, u-n) \delta^{(n)}(\mathbf{x}) = 
u \delta^{(n)}(\mathbf{x}), $
and from these the trivial product N-point YSC,
$$ \Phi_{K,N}(\mathbf{x}_1,...\mathbf{x}_N; \mathbf{u}) = 
\prod_1^K  \delta^{(n)}(\mathbf{x}_i), \ \ 
 E (\mathbf{u}) = u_1...u_K (u_{K+1}+1) ...(u_N+1), $$
$$
 \mathbf{u} = 
\begin{pmatrix}
u_1 & ...& u_K& u_{K+1}& ...& u_N \\
u_1-n & ...&u_K-n&u_{K+1}&...& u_N 
\end{pmatrix}.
$$

The Yang-Baxter $RLL$ relations provide a way to non-trivial YSC.
We have

$$ R_{12}(u_1-u_2) \ L_1(u_1^+, u_1)L_2(u_2^+,u_2) = 
L_1(u_1^+, u_2)L_2(u_2^+,u_1) \ R_{12}(u_1-u_2),
$$

$$ R_{21}(u_1^+-u_2^+) \ L_1(u_1^+,u_1)L_2(u_2^+,u_2) = 
L_1(u_2^+,u_1)L_2(u_1^+,u_2) \ R_{21}(u_1^+-u_2^+),
$$
where the $R$ operator can be represented by
\be \label{R12} 
R_{12}(u) = \int \frac{dc}{c^{1+u}} 
e^{-c (\mathbf{x}_1 \cdot \mathbf{p}_2)} \ee
with the integration over a closed contour. 
If $\Phi(\mathbf{u})$ is a YSC obeying (\ref{YSC} then
$\Phi(\mathbf{u}\p ) = R_{i, i+1} (u_i-u_{i+1} )\Phi(\mathbf{u})$ 
obey the YSC relation with $\mathbf{u}$ replaced by

$$ \mathbf{u}\p = 
\begin{pmatrix}
u_1 & ...& u_{i+1}& u_{i}& ...& u_N \\
u_1^+ & ...& u_i^+& u_{i+1}^+&...& u_N 
\end{pmatrix},
$$
i.e. the entries at $i, i+1 $ in the first row are permuted. 
$\Phi(\mathbf{u}\pp ) = R_{ i+1,i} (u_i^+-u_{i+1}^+) \Phi(\mathbf{u})$ 
obeys the YSC relation with  $\mathbf{u}$ replaced by

$$\mathbf{u}\pp = 
\begin{pmatrix}
u_1 & ...& u_{i}& u_{i+1}& ...& u_N \\
u_1^+ & ...&u_{i+1}^+&u_{i}^+ & ...& u_N 
\end{pmatrix},
$$

i.e. the entries at $i, i+1 $ in the second row are permuted. 
The resulting YSC are less trivial and by repeated $R$ operations we 
obtain completely connected correlators characterized by the permuted 
parameter set of the resulting monodromy matrix, 

\be \label{bfu}
 \begin{pmatrix}  
u_{\sigma(1)} & u_{\sigma(2)} & \dots & u_{\sigma(N)} \\
u^+_{\bar \sigma(1)}& u^+_{\bar \sigma(2)} & \dots & u^+_{\bar \sigma(N)} 
\end{pmatrix}.
\ee

We shall use  the abbreviation  of writing the indices carried by the
parameters only. Notice that by the substitution of parameters 
$u_{\sigma(1)}, u_{\sigma(2)}, \dots  , u_{\sigma(N)} \rightarrow 
v_1, v_2, \dots ,v_N
$ 
applied to both rows the first row can be put into the standard ordering.

One may draw an analogy of (\ref{YSC}) to the time-independent Schr\"odinger
equation. In this respect we have considered so far the position 
 representation. Besides of that we need the helicity
representation. It is defined in the case of even $n, n= 2m$  
by the elementary canonical transformation applied to half of the
canonical pairs at each point, $a= m+ \alpha = m+1, ..., n$, 
leaving the first half
$ a = \alphadot, \alphadot = 1, ..., m$ unchanged, 
\be \label{canon1}
 \begin{pmatrix}
x_{i,m+\alpha} \\ \dd_{i,m+\alpha} 
\end{pmatrix} \rightarrow
 \begin{pmatrix}
\dd^{\lambda}_{i,\alpha} \\ -\lambda_{i, \alpha} 
\end{pmatrix},  \ \ \ 
\begin{pmatrix}
x_{i,\alphadot} \\ \dd_{i, \alphadot} 
\end{pmatrix} \rightarrow
\begin{pmatrix}
\bar \lambda_{i,\alphadot} \\ \bar \dd^{\lambda}_{i,\alphadot} 
\end{pmatrix}. 
\ee

The $N$-point YSC can be written in the link integral form  
which became a standard of the modern tools of amplitude calculations
\cite{AH}.
In the position representation it is given by
\be \label{Phipos}
 \Phi_{K,N} (\mathbf{x}_1,...\mathbf{x}_N; \mathbf{u})  =
\int d c \varphi (\mathbf{c}) \prod_{i=1}^K \delta^{(n)}  ( \mathbf{x}_i -
\sum_{j= K+1}^N c_{ij} \mathbf{x}_j ). \ee
$K$ denotes the number of $\delta^{(n)}$ factors.  

To change to the helicity representation we apply the Fourier transformation
to the dependence on the components $x_{i,a}, a=m+1, ...,n$ at each point
$i=1, ...,N$. The Fourier variable conjugate to $x_{m+\alpha}$ is denoted by
$\lambda_{i,\alpha}, \alpha = 1, ...,m$. 
The components $x_{i,a}, a= 1, ...,m$ are not changed and merely renamed by
$\bar \lambda_{i,\dot \alpha}$, i.e. 
$\bar \lambda_{i,\dot \alpha}  = x_{i,\dot \alpha}, \dot \alpha = 1, ...,m$.
Thus the link integral form of a YSC in 
 the helicity representation is
\be \label{Phihel}
 \Phi^{\lambda}_{K,N} (\bar \lambda_1, \lambda_1, ...\bar \lambda_N, \lambda_N; 
\mathbf{u})  =
\int d c \varphi (\mathbf{c}) 
\prod_{i=1}^K \delta^{(m)} ( \bar \lambda_i -
\sum_{j= K+1}^N c_{ij} \bar \lambda_j ) 
\prod_{j=K+1}^N \delta^{(m)} (  \lambda_j +
\sum_{i= 1}^K c_{ij}  \lambda_i ). 
\ee
Notice that the integrand function $\varphi$ is the same in both
representations. $d c$ abbreviates $d c = \prod_{i=1}^K \prod_{j=K+1}^N
dc_{ij} $.

\section{Examples of YSC for N=2, 3, 4 }
 \setcounter{equation}{0}

We present examples of YSC encountered in the applications below and
indicate their construction by $R$ operations. 

We start with the two-point YSC, $N=2, K=1$
\be \label{phi12}
\Phi_{1,2}(\mathbf{x}_1, \mathbf{x}_2;\mathbf{u}) = 
R_{21} (u_1^+ - u_2) \delta^{(n)} (\mathbf{x}_1), \ee
$$\varphi_{1,2}(c_{12} )  = c_{12}^{-1+ u_2 +n-u_1}, \ \ \ 
\mathbf{u} = 
\begin{pmatrix}
1 & 2 \\
2 & 1^+
\end{pmatrix}. 
$$
In the second row $2$ stands for $ u_2$ and $1^+$ for $u_1-n= u_1^+$. 

Next we give the three-point YSC with $N=3, K=1$
\be \label{phi13}
\Phi_{1,3}(\mathbf{x}_1, \mathbf{x}_2,\mathbf{x}_3;\mathbf{u}) =
 R_{21}^{++}(u^+_3-u^+_2) R_{31}^{++}(u_1^+-u_3^+) \cdot
\delta^{(n)}(\mathbf{x}_1 ), \ee
 $$ \varphi_{1,3} = \left ( c_{12}^{1+ u_3 - u_2} c_{13}^{1+ u_1^+ - u_3}
\right)^{-1}, \ \ \ 
 \mathbf{u} = 
\begin{pmatrix}
1&2&3 \\
2&3&1^+
\end{pmatrix},
$$
and with $N=3, K=2$,
\be \label{phi23}
 \Phi_{2,3} (\mathbf{u}) = R_{32} (u_2^+ - u_1^+) R_{31}(u_1^+ - u_3^+)
\delta^{(n)}(\mathbf{x}_1 ) \delta^{(n)}(\mathbf{x}_2 ), \ee
 $$ \varphi_{2,3} = \left (c_{23}^{1+u_2^+ - u_1^+} c_{13}^{1+ u_1^+ - u_3^+}
\right )^{-1}, \ \ \ 
 \mathbf{u} = 
\begin{pmatrix}
1&2&3 \\
3&2^+&1^+
\end{pmatrix}.
$$

We give also the relevant examples of four-point YSC. 

As an intermediate YSC from which the following ones are constructed we
write

\be \label{phiDelta}
 \Phi_{\Delta} (\mathbf{u}) = R_{32} (u_2^+ - u_3^+) R_{41} u_1^+ -u_4^+)
\delta^{(n)}(\mathbf{x}_1 ) \delta^{(n)}(\mathbf{x}_2 ), \ee 
$$ \varphi_{\Delta} = \delta(c_{13}) \delta(c_{24}) \left ( c_{23}^{1+ u_2^+ -
u_3^+} c_{14}^{1+ u_1^+ - u_4} \right )^{-1}, 
\ \ \  \mathbf{u} = 
\begin{pmatrix}
1&2&3&4 \\
4&3&2^+& 1^+
\end{pmatrix}.
$$
This YSC is incompletely connected with delta distributions in $\varphi$.
 It takes two more steps of $R$ operations
to arrive at the completely connected one.

\be \label{phix}
 \Phi_{X} (\mathbf{u}) =  R_{43}(u_2^+ - u_4^+) R_{21}(u_4^+ - u_3^+)
\Phi_{\Delta}, \ee 
$$
\varphi_{X} = \left ( c_{24}^{1+u_2-u_1} c_{13}^{1+u_4-u_3} (c_{14} c_{23} -
c_{23} c_{13})^{1+u_2^+ -u_3} \right )^{-1}, $$
$$ \mathbf{u} = 
\begin{pmatrix}
1&2&3&4 \\
3&4&1^+& 2^+
\end{pmatrix}.
$$
The sequence of operations leading to a YSC is not unique. In this case we 
may also choose
$$ \Phi_{X} (\mathbf{u}) =  R_{12}(u_1 - u_2) R_{21}(u_4^+ - u_3^+)
\Phi_{\Delta} |_{u_1 \leftrightarrow u_2} $$
 with the indicated substitution as the last step to obtain the standard
ordering in the upper row of $u$ parameters.

The permutation pattern $\mathbf{u}$ fixes the monodromy operator. However,
the YSC is not uniquely fixed. For example, the  pattern
$$ \mathbf{u} = 
\begin{pmatrix}
1&2&3&4 \\
4&3&2^+& 1^+
\end{pmatrix}
$$
is the same as for $\Phi_{\Delta}$ and is also the one of the following two
YSC.
$$ \Phi_{||1} (\mathbf{u}) = R_{34}(u_3-u_4) R_{12} (u_1-u_2)
R_{12}(u_1 - u_2) R_{21}(u_4^+ - u_3^+)
\Phi_{\Delta} |_{u_3 \leftrightarrow u_4}, $$
$$\varphi_{||1} = \left ( c_{24} c_{13} c_{14}^{  u_1-u_2+u_3-u_4}  
 (c_{14} c_{23} - c_{23} c_{13})^{1+u_2^+ -u_3} \right )^{-1}, $$

$$ \Phi_{||2} (\mathbf{u}) = R_{43}(u_2^+-u_1^+) R_{21} (u_3^+-u_4^+)
R_{12}(u_1 - u_2) R_{21}(u_4^+ - u_3^+)
\Phi_{\Delta} |_{u_3 \leftrightarrow u_4}, $$
$$\varphi_{||2} = \left ( c_{24} c_{13} c_{23}^{ 1+ u_2-u_1+u_4-u_3}  
 (c_{14} c_{23} - c_{23} c_{13})^{1+u_1^+ -u_4} \right )^{-1}. $$
There are more YCS with the above parameter pattern. In the following we
shall use
$ \Phi_{||} (\mathbf{u}) $ with
\be \label{phi=}
\varphi_{||} = \left ( c_{24} c_{13} c_{14}^{ u_3-u_4} c_{23}^{u_2-u_1}  
 (c_{14} c_{23} - c_{23} c_{13})^{1+u_1^+ -u_3} \right )^{-1}. \ee
It is the result of the convolution of two YSC of the form $\Phi_X$.
At $u_1=u_2$ it coincides with $\Phi_{||1}$ and at $u_3=u_4$ it coincides
with $\Phi_{||2}$.

The integration variables $c_{ij} $ may be regarded as coordinates on the
corresponding Grassmannian variety $\mathcal{G}_{K,N}$ and in the case of
completely connected YSC the closed
integration contours extend over a maximal Schubert cell.
The delta distributions involving the correlator variables $\mathbf{x}$ or
$\bar \lambda, \lambda$ fix some of these integration variables $c_{ij} $. 
In the following applications we have cases where all are fixed,
e.g. in the position representation for $n=2$  with  $(K,N)= (1,3) $ and 
$(K,N)= (2,4) $. The helicity representation has the feature that
the delta distributions can be rewritten in a way with the
factor $\delta(\sum_1^N \bar \lambda_i \lambda_i) $ independent of $c_{ij}$.
The integrations are removed in this representation
for $n=4$ with $(K,N) = (2,4)$.

\section{The BFKL kernel}
\label{BFKL}
 \setcounter{equation}{0}

YSC may be used as kernels of integral operators. The symmetry of the
kernels implies symmetry properties like Yang-Baxter relations for these
operators. At $n=2$ we may change the homogeneous coordinates 
$\mathbf{x}_i  = (x_{i,1}, x_{i,2})$ to
the normal coordinates $x_i = \frac{x_{i,1}}{x_{i,2}} $ and find
$$ \Phi (\mathbf{x}_1, ..., \mathbf{x}_N) = \prod_1^N x_{i,2}^{2\ell_i}
\phi(x_1, ..., x_N). $$   

We  define the integral operator with a 4-point
YSC as kernel,
$$ [R \psi](x_1,x_2) = \int dx_1\p dx_2\p \psi(x_1\p, x_2\p)
\phi(x_1\p, x_2\p, x_1,x_2|\mathbf{u}). 
$$
In the  case of the action on functions on the complex plane we consider
the complex $x$ as $(x, \bar x)$ and define
$$ [R \psi](x_1,x_2) = \int dx_1\p dx_2\p d\bar x_1\p d\bar x_2\p
\psi(x_1\p, x_2\p,\bar x_1\p, \bar x_2\p)
\phi(x_1\p, x_2\p, x_1,x_2|\mathbf{u})\phi(\bar x_1\p, \bar x_2\p, \bar x_1,
\bar x_2|\bar {\mathbf{u}}).
$$
We substitute the YSC at $n=2$ in the normal coordinate form
and perform the  integrals over $c_{13}, c_{14}, c_{23}, c_{24}$.
$$ \phi(x_1\p, x_2\p, x_1,x_2|\mathbf{u}) = x_{12}^{-2} \varphi(c^*). $$

Provided the substitution $1,2,3,4,\to 1\p ,2\p,1,2$ we have
$$ c^*_{13} =  \frac{x_{14}}{x_{34}}, \ \  
c^*_{14} =  -\frac{x_{13}}{x_{34}}, \ \ 
c^*_{23} =  \frac{x_{24}}{x_{34}}, \ \ 
c^*_{24} =  -\frac{x_{23}}{x_{34}}.
$$
We use the abbreviation $x_{ij} = x_i-x_j$ for the difference of normal
coordinates (not to be mixed with the component notation $x_{i,1},
x_{i,2}$).

In the case of the  4-point YSC (\ref{phix}) 
we obtain
$$\phi_X (x_1,x_2,x_3,x_4)= \frac{x_{34}^{1+u_2^+ -u_3} }{ x_{23}^{1+u_2-u_1}
x_{14}^{1+u_4-u_3} x_{12}^{1+u_1^+-u_4}} =
\frac{x_{12}^{1+2\ell_1-\e} }{x_{23}^{1+2\ell_1-2\ell_2-\e} x_{14}^{1-\e}
x_{24}^{1+2\ell_2+ \e} }. $$
In the last step we have used the relation between the spectral parameters
and the weight at each point $i$, $2\ell_i = u^+_{\bar \sigma(i)} -
u_{\sigma(i)} $, referring to the parameter permutation pattern (\ref{bfu}).
In our case (\ref{phix}) we have
$$2\ell_1 = u_3-u_1, 2\ell_2 = u_4-u_1, 2\ell_3 = u_1-n-u_3= -2\ell_1-n, 
2\ell_4= u_2-n-u_4 = -2\ell_2-n $$
We introduce $\e= u_3-u_4$ and express the exponents in terms of the
independent weights $2\ell_1, 2\ell_2$ and the parameter $\e$, substituting
also $n=2$ for the considered case.

We substitute $1,2,3,4,\to 1\p ,2\p,1,2$ .
At $\ell_{1\p}=\ell_{2\p}=\ell$ the kernel for the complex plane is

$$ \phi(x_1\p, x_2\p, x_1,x_2)\phi(\bar x_1\p, \bar x_2\p, \bar x_1,
\bar x_2) =
 \frac{|x_{1\p 2\p}|^{2(-1-2\ell-\e)} |x_{1 2}|^{2(1+2\ell-\e)} }{
|x_{1\p 2}|^{2(1-\e)} |x_{1 2\p}|^{2(1-\e)} }. 
$$
In the decomposition in $\e$ the leading $\e^{-1}$ term corresponds to the
kernel of the 
permutation operator $P_{12}$. The finite term is
$$|x_{1\p 2\p}|^{2(-1-2\ell)} |x_{1 2}|^{2(1+2\ell)} 
\left ( |x_{1\p 2}|^{2} \delta^{(2)} (x_{1 2\p})  +|x_{1 2\p}|^{2} 
\delta^{(2)} (x_{1\p 2}) - 
\delta^{(2)} (x_{1 2\p})\delta^{(2)} (x_{1\p 2}) 
\int d^2 x_3 |x_{1 3}|^{2} |x_{1 3}|^{2} \right ) .
$$
 The subtraction is the appropriately modified $+$ prescription.
In the limiting case corresponding to the reggeized gluon exchange 
$\ell = 0$ the action is defined on functions
vanishing at coinciding arguments $x_{1\p} = x_{2\p}$. The kernel
 can be better rewritten as
$$  |x_{1 2}|^{2}\int \frac{d^2 x_3}{|x_{1 3}|^{2}|x_{23}|^{2} }
 \left ( \delta^{(2)} (x_{2 1\p}) \delta (x_{2\p 3})  
+ \delta^{(2)} (x_{1 2\p}) \delta (x_{1\p 3})
- \delta^{(2)} (x_{1 2\p}) \delta (x_{21\p} )
\right ). 
$$
This is the dipole (or Moebius) form of the BFKL kernel
\cite{YB99,YK99,BLV04, RK05}.

The other argument line to BFKL starts with noticing that the
integral operator obeys the standard RLL relation for intertwining
representations with weights $\ell_1, \ell_2$.
 The $x ,\dd$ operator form of its holomorphic part can be represented in
the factorized form \cite{SD05} , $R(\e) = R^1(\e)  R^2(\e) $,
$$ R^1(u^1|v^1, v^2) = \frac{\Gamma(x_{21} \dd_2 + \e + \ell_1 +\ell_2 +1)}{ \Gamma(x_{21} 
\dd_2 +
2\ell_2 +1) }, \ \ \  
 R^2 (u^1,u^2|v^2)=  \frac{\Gamma(x_{12} \dd_1 + \e + \ell_1 +\ell_2 +1)}{ \Gamma(x_{12} 
\dd_1 +2\ell_1 +1) }, $$
and then the decomposition in $\e$ results in one of the
operator forms of the BFKL operator (compare e.g. \cite{Levrev96}, eq.184).
This line of arguments is the one in the original  paper by L.N. Lipatov
on the relation to the integrable spin chains \cite{LNL93}.

\section{The parton evolution kernels}
\label{DGLAP} 
\setcounter{equation}{0}

In the helicity representation we have at $n=2, m=1$,
$$\Phi( \bar \lambda_1, \lambda_1, \dots, \bar \lambda_N, \lambda_N) =
\prod_1^N \bar \lambda_i^{2h_i} \ \delta (\sum k_i)
\phi (k_1, ..., k_N), $$
where $k_i = \bar \lambda_i \lambda_i $  
 have one component only in this case.
In the application to parton evolution these variables  $k_i$ have the physical
meaning of light-cone momenta and the arguments $z$ of the parton
evolution kernels are defined as their ratios.
We obtain for the YSC with $N=3$ and $K=1$ (\ref{phi13}) or $K=2$ (\ref{phi23}) 
$$
\phi(k_1,k_2,k_3; a_1,a_2,a_3) = (k_1 k_2 k_3)^{\half} k_1^{- \eta a_1}
k_2^{- \eta a_2} k_3^{- \eta a_3}. 
$$
Here $2a_i = 2\ell_1 + 1 $ and $ \sum_1^3 2 a_i = \eta $, with
$\eta= \pm 1$. The positive sign $\eta=+1$ corresponds to $K=1$ (\ref{phi13}) 
and the negative $\eta = -1$ to $K=2$ (\ref{phi23} ).
 The amplitude of parton splitting $3 \to 1+2 $
is obtained from this three-point YSC by substitutions of
$k_i$ in terms of the momentum fraction  $z$ and $a_i$ 
by the  parton helicities $h_i$ as follows
$$ 
Split(h_1, h_2,h_3;z) = \phi( -z,  z-1, 1; h_1,h_2, h_3- \half \eta ). 
$$

The parton splitting probabilities are calculated as squares of the
corresponding splitting amplitudes. The helicities $h_i$ 
refer to ingoing momenta,
i.e. $h_1,h_2$ are opposite to their physical values in the decay
$3 \to 1+2$.
$$ P _{ h_1 \ h_3}^{h_2} (z) = 
\left ( Split(h_1,h_2,h_3;z) \right )^2 = 
\left (\phi(-z,-1+z, 1; h_1,h_2, h_3- \half \eta ) \right )^2. 
$$
The expressions for the leading order  parton evolution kernels 
are reproduced, compare e.g.\cite{Levrev96}.

\section{Scattering amplitudes}
\label{Ampl} 
\setcounter{equation}{0}

We consider the four-point YSC at $n=4$ with $K=2$ in the helicity
representation and do the integrals over
the $c$ variables, 
\be \label{philambda}
 \Phi_{2,4}^{\lambda} = \varphi(c^*) \ \ \delta^{(4)}(\sum k_i), \ee
$$ c_{13}^* = \frac{[14]}{[34]}, \  
 c_{14}^* = \frac{[13]}{[43]}, \ 
 c_{23}^* = \frac{[24]}{[34]}, \   
 c_{24}^* = \frac{[23]}{[43]}. $$  
Here we denote $ (k_i)_{\alphadot, \alpha} = 
\bar \lambda_{i,\alphadot} \lambda_{i,\alpha}$,   
$[ij] = \bar \lambda_{i,1} \bar \lambda_{j,2} - \bar \lambda_{i,2} \bar
\lambda_{j,1}$.

From (\ref{philambda}) we obtain the explicit expressions as functions of the
independent helicities and the extra parameter $\e$.
As in sect. 4 we use the relation between the spectral parameters and the
weights referring to the parameter permutation scheme $\mathbf{u}$
 (\ref{bfu}). The
weights are substituted by the helicities as $2\ell_i +2 = 2 h_i$. In the
case (\ref{phix}) we have 
$$2h_1 = u_3-u_1+2 = -2h_3, 2h_2 = u_4-u_2+2 = -2h_4. $$
We introduce $\e = u_3-u_4$ and express the exponents in terms of the
helicities $h_1, h_2$ and the parameter $\e$.
$$ \varphi_{X}(c^*; h_1,h_2, \e) = 
\left ( \frac{ [14] [23]}{[12][34]} \right )^{\e} \
\frac{[12]^{1+2h_1} [34]^{1-2h_2} }{ [14] [23]^{1+2h_1-2h_2} }.
$$ 
Thus in the case of $\Phi_X$ the helicities are related by
$ h_1 = - h_3, \ \ h_2 = -h_4 $.
With these conditions only 4 of the 6 helicity configurations of the
$2 \to 2$ helicity amplitudes are accessible.

In the case (\ref{phi=}) we have instead
$$2h_1 = u_4-u_1+2= - 2h_4, 2h_2 = u_3-u_2+2 = -2h_3. $$
We introduce $\e= u_4-u_3$ and express the exponents in terms of the
helicities $h_1,h_2$ and the parameter $\e$. 
$$ \varphi_{||}(c^*; h_1,h_2, \e) = 
\left ( \frac{ [13] [24]}{[12][34]} \right )^{\e} \
\frac{[12]^{1+2h_1} [34]^{1-2h_2} }{ [14] [23] [24]^{2h_1-2h_2} }.
$$ 
With $\Phi_{||}$ we
 cover the remaining cases because here the helicities are
related by
$ h_1 = - h_4, \ \ h_2 = -h_3 $.
With these conditions again 4 helicity configuration cases can be covered,
two of them doubling some of the above correlator.

We observe that the gluon tree amplitudes in spinor-helicity form
\cite{Dixon96,Khoze04} are reproduced by these two
YSC as

$$ M(1,1,-1,-1) = \varphi_{X}(c^*, 1,1,0), \ \    
M(-1,-1,1,1) = \varphi_{X}(c^*, -1,-1,0), \ \
$$ $$
 M(1,-1,-1,1) = \varphi_{X}(c^*, 1,-1,4), 
 M(-1,1,1,-1) = \varphi_{X}(c^*, -1,1,0), $$

$$ M(1,1,-1,-1) = \varphi_{||}(c^*, 1,1,0), \ \    
M(-1,-1,1,1) = \varphi_{||}(c^*, -1,-1,0), \ \
$$ $$ 
M(1,-1,1,-1) = \varphi_{||}(c^*, 1,-1,4), 
 M(-1,1,-1,1) = \varphi_{||}(c^*, -1,1,0).
$$

The quark tree amplitudes are reproduced as

$$
M(\half, \half,-\half ,-\half) = \varphi_X(c^*, \half,\half, 0), $$ $$ 
M(\half, -\half,-\half, \half) = \varphi_X(c^*, \half,-\half, 3), \ \ 
M(-\half, \half,\half, -\half) = \varphi_X(c^*, -\half,\half, 1), \ \ 
$$
$$
M(\half, \half,-\half ,-\half) = \varphi_{||}(c^*, \half,\half, 0), $$ $$ 
M(\half, -\half,\half, -\half) = \varphi_{||}(c^*, \half,-\half, 3), \ \ 
M(-\half, \half,-\half, \half) = \varphi_{||}(c^*, -\half,\half, 1).  
$$

The gluon-quark tree amplitudes are reproduced as

$$ M(1,\half, -1,-\half) = \varphi_X(c^*, 1, \half, 1), \ \ 
M(1,-\half, -1,\half) = \varphi_X(c^*, 1, -\half, 3),
$$
$$
M(1,\half, -\half,-1) = \varphi_{||}(c^*, 1, \half, 1), \ \ 
M(1,-\half, \half,-1) = \varphi_{||}(c^*, 1, -\half, 3).
$$

\section{Asymptotic freedom}
\label{GML} 
\setcounter{equation}{0}

We reconsider the convolution of 3-point YSC with a two-particle
intermediate state.

$$
\int d^n \mathbf{x}_{1\p} d^n \mathbf{x}_{2\p} \Phi_{1,3}(1,2\p,1\p) \ 
\Phi_{2,3}(1\p,2\p,2) = $$ $$
\int dc_{1,1\p} dc_{1 2\p} \varphi_{1,3}(1,2\p,1\p) dc_{2\p 2} dc_{1\p 3}
\varphi_{2,3}(1\p,2\p,2)
\delta(\mathbf{x}_1 - c_{1 1\p} \mathbf{x}_{1\p} - c_{1 2\p}
\mathbf{x}_{2\p} ) \ 
\delta(\mathbf{x}_{1\p} - c_{1\p 2} \mathbf{x}_2 )
\delta(\mathbf{x}_{2\p} - c_{2\p 2} \mathbf{x}_2 ) $$

In the function arguments we have abbreviated the points $\mathbf{x}_i$
by the indices $i$.
The result can be written in the form of a two-point correlator
$$
\int d\bar c_{12} \bar \varphi(\bar c) 
\delta(\mathbf{x}_{1} - \bar c_{1 2} \mathbf{x}_2 ),
 $$
where
$$ \bar \varphi(\bar c) =
\int \frac{dc_{2\p 2}}{c_{2\p 2}} dc_{1 1\p} dc_{1\p 2} 
\varphi_{1,3}(\frac{\bar c_{12} - c_{1 1\p}c_{1\p 2} }{ c_{2\p 3} }, c_{1
1\p}) \varphi_{2,3} ( c_{2\p,2}, c_{1\p,2}). $$

We substitute (\ref{phi13}) and (\ref{phi23}) substituting the variables
correspondingly.
The spectral parameters of the first factor are denoted by 
$u_1, u_{2\p},u_{1\p} $ and the ones of the second factor by $v_{1\p},
v_{2\p}, v_2$.

$$  \bar \varphi(\bar c) =\int \frac{dc_{2\p 2}}{c_{2\p 2}} dc_{1 1\p}
dc_{1\p 2} 
\left ( (\bar c_{12} - c_{1 1\p}c_{1\p 2})^{1+ u_{1\p}-u_{2\p}} c_{11\p}^{1+u_1^+
-u_{1\p} } c_{2\p 2}^{1+ v_{2\p}^+ - v_{1\p} + u_{2\p} - u_{1\p}} 
c_{1\p 2}^{1+ v_{1\p}^+ - v_2} 
\right )^{-1}
$$

We calculate the weights in terms of the
spectral parameters marking the weights of the second factor by a prime.
$$ 2\ell_1 = u_{2\p} - u_1, 2\ell_{2\p} = u_{1\p} - u_{2\p}, 2\ell_{1\p} =
u_1^+ - u_{1\p}, $$
$$ 2\ell\p_{1\p} = v_2 -v_{1\p}, 2\ell\p_{2\p} = v_{1\p}^+ - v_{2\p},
2\ell\p_2 = v_{2\p} -v_2 . $$
We impose the weight balance condition, 

$$ v_{2\p}^+ - v_{1\p}^+ = -2\ell\p_{2\p} -n = 2\ell_{2\p},
v_{1\p}^+ -v_2 = -2\ell\p_{1\p} -n = 2\ell_{1\p}. $$
By the change of integration variables
$ c_{11\p} \to c = \frac{c_{11\p}c_{1\p 2}}{ \bar c_{12}} $
we obtain

$$    
\bar \varphi(\bar c) = 
\int \frac{dc_{2\p 2} dc_{1\p 2} }{ c_{2\p 2}c_{1\p 2} }  
 \ \ B(-2\ell_{1\p}, -2\ell_{2\p} ) 
\frac{1}{\bar c_{13}^{1+2\ell_{1\p} + 2 \ell_{2\p} } } 
$$

Thus the result of the symmetric convolution of three-point YSC is the
 two-point YSC multiplied by a factor depending on the weights of the
intermediate two-particle state,
$$
\int d \mathbf{x}_{1\p} d \mathbf{x}_{2\p} \Phi_{1,3}(1,2\p,1\p) \ 
\Phi_{2,3}(1\p,2\p,2) =  
const \ B(-2\ell_{1\p}, -2\ell_{2\p}) \Phi_{1,2}(1,2) $$

We study the resulting dependence on the intermediate state
weights entering the Euler beta function factor. We have 
$2\ell_1 + 2\ell_{1\p} + 2\ell_{2\p} + n = 0,$  and we change to the
helicities $ 2\ell + \half n = 2 h $.
$$ B(- 2\ell_{1\p}, -2\ell_{2\p}) = B(2h_1 + 2h_{2\p}, -2h_{2\p}   +\half n)
$$
In the interesting case of QCD, where  
$ n=4, h_1 = 1 $, we have with 
$2h_{2\p} \to 2h = 2\bar h + 2\e$, $2 \bar h$ integer, 

$$ B(2h+2, -2h+2) 
- \frac{1}{12 \e}  (2\bar h+1) 2\bar h (2\bar h-1) +
(-1)^{2\bar h} \frac{-12 \bar h^2 +1}{ 6 } + \mathcal{O}(\e). $$

The finite term results in the contributions to the leading Gell-Mann -Low
coefficient with the values $\bar h=1$ for the gluon loop contribution and 
$\bar h= \half $ for 
the quark loop contribution.

\section*{Acknowledgements}
{ This work has been supported in part by JINR Dubna in the frame of the
Heisenberg-Landau program.}

\end{document}